\documentclass{ws-ijmpb}
\def\ket#1{| #1 \rangle}
\def\bra#1{\langle #1 |}
\begin{document}

\markboth{A.I. Solomon and Sonia G. Schirmer}{Limitations on Quantum Control}
\catchline{}{}{}

\title{LIMITATIONS ON QUANTUM CONTROL}

\author{\footnotesize ALLAN I. SOLOMON\footnote{Permanent address: Quantum Processes Group, Open University, Milton Keynes MK7 6AA, United Kingdom. Email: a.i.solomon@open.ac.uk}}
\address{Laboratoire de Physique Th\'eorique des Liquides, University of Paris VI, France
}
\author{SONIA G. SCHIRMER\footnote{Email: s.g.schirmer@open.ac.uk}}
\address{Quantum Processes Group, Open University\\
Milton Keynes MK7 6AA, United Kingdom}
\maketitle

\pub{Received (received date)}{Revised (revised date)}

\begin{abstract}
In this note we give an introduction to the topic of Quantum Control, 
explaining what its objectives are, and describing some of its limitations.
\end{abstract}

\section{What is Quantum Control?}

The objectives of Quantum Control are to find ways to manipulate the
time evolution of a quantum system such as to
\begin{itemlist}
\item drive an initial given state to a pre-determined final state, 
      the target state; or
\item optimize the expectation value of a target observable.
\end{itemlist}

Among a wealth of  applications are those to Quantum Computing, where it is clearly 
essential to be able to start off a quantum procedure with a given initial state, and
to problems involving the population levels in atomic systems, such as the laser cooling
of atomic or molecular systems.

The mathematical tools necessary for the theoretical investigation of these control 
problems are diverse, involving algebraic, group theoretic and topological methods.

The questions  that one may ask include:
\begin{romanlist}
\item  When is a given quantum system completely controllable?
\item  If a system is not completely controllable, how does this affect optimization 
       of a given operator?
\item  How near can you get to a target state for a not completely controllable system?
\end{romanlist}
The answer to the first question depends on a knowledge of the Lie algebra generated by
the system's quantum hamiltonian, that to the second arises from properties of the Lie
group structure, while the last clearly involves ideas of topology.

Especially in the area of Lie group theory, there is a large corpus of classical 
mathematics which can supply answers to questions arising in quantum control.  In 
particular, for the type of controllability known as {\em Pure State Controllability}
classical Lie Group theory has already given the basic results.

The quantum control system we shall consider is  typically of the form
\begin{equation} \label{eq:H}
  {H} = {H}_0 + \sum_{m=1}^M f_m(t) {H}_m,
\end{equation}
where ${H}_0$ is the internal Hamiltonian of the unperturbed system and ${H}_m$ are 
interaction terms governing the interaction of the system with an external field.  
The dynamical evolution of the system is governed by the unitary evolution operator 
$ {U}(t,0)$, which satisfies the Schrodinger equation
\begin{equation} \label{eq:SE}
 i\hbar \frac{\partial}{\partial t} {U}(t,0) =  {H}  {U}(t,0)
\end{equation}
with initial condition ${U}(0,0)= {I}$, where $ {I}$ is the identity operator.  By 
use of the Magnus expansion, it can be shown that the solution $U$ involves all the 
commutators of the $ {H}_m$.  

The operators $ {H}_m$, $0\le m\le M$, in (\ref{eq:H}) are Hermitian.  Their 
skew-Hermitian counterparts ${\rm i} {H}_m$ generate a Lie algebra $L$ known as the 
dynamical Lie algebra of the control system which is always a subalgebra of $u(N)$,
or for trace-zero hamiltonians, $su(N)$.  The degree of controllability is determined
by the dynamical Lie algebra generated by the control system hamiltonian ${H}$.  If 
$L=u(N)$ then {\em all} the unitary operators are generated and we call such a system 
{\em Completely Controllable}.  A large variety of common quantum systems  can be shown
to be Completely Controllable \cite{us1,us2}.  The interesting cases arise when  $L$ 
is a proper subalgebra of $u(N)$.  Such systems may still exhibit {\em Pure State 
Controllability}, in that starting with any initial pure state any target pure state 
may be obtained, as distinct from the Completely Controllable case, when all 
(kinematically admissible) states --- pure or mixed --- may be achieved.

\section{Pure state controllability}

We shall restrict our attention here to finite-level quantum systems with $N$ discrete 
energy levels.  The pure quantum states of the system are represented by normalized
wavefunctions $\ket{\Psi}$, which form a Hilbert space $H$.  However, the state of a
quantum system need not be represented by a pure state $\ket{\Psi}\in H$.  For instance,
we may consider a system consisting of a large number of identical, non-interacting
particles, which can be in different internal quantum states, i.e., a certain fraction 
$w_1$ of the particles may be in quantum state $\ket{\Psi_1}$, another fraction $w_2$
may be in another state $\ket{\Psi_2}$ and so forth.  Hence, the state of the system as
a whole is described by a discrete ensemble of quantum states $\ket{\Psi_n}$ with
non-negative weights $w_n$ that sum up to one.  Such an ensemble of quantum states is
called  a {\em mixed-state}, and it  can be represented by a density operator $\rho_0$
on $H$ with the spectral representation 
\begin{equation} \label{eq:rho0}
   {\rho}_0 = \sum_{n=1}^N w_n \ket{\Psi_n} \bra{\Psi_n},
\end{equation}
where $\{\ket{\Psi_n}: 1\le n\le N\}$ is an orthonormal set of vectors in $H$ that
forms an ensemble of independent pure quantum states.  The evolution of $ {\rho}_0$ 
is governed by
\begin{equation} \label{eq:rhot}
   \rho(t) =  U(t,0) \rho_0 U(t,0)^\dagger,
\end{equation}
with $U(t,0)$ as above.  Clearly if {\em all} the unitary operators can be generated
we have the optimal situation, complete controllability.  However, classical Lie group
theory tells us that even when we only obtain a subalgebra of $u(N)$ we can obtain 
pure state controllability.

The results arise from consideration\cite{mont} of the transitive action of Lie groups
on the sphere $S^k$.  The classical ``orthogonal'' groups $\Theta(n,F)$ where the field
$F$ is either the reals $\Re$, the complexes ${\cal C}$ or the quaternions ${\cal H}$, 
are defined to be those that keep invariant the length of the vector $v\equiv (v_1,v_2,
\ldots,v_n)$; the squared length is given by $v^\dagger v=\sum_{i=1}^n{\bar{v_i} v_i}$,
where $\bar{v}_i$ refers to the appropriate conjugation.  These compact groups are, 
essentially, the only ones which give transitive actions on the appropriate spheres, 
as follows:
\begin{romanlist}
\item $\Theta(n,\Re) \equiv O(n)$ transitive on $S^{(n-1)}$
\item $\Theta(n,{\cal C}) \equiv U(n)$ transitive on $S^{(2n-1)}$ 
\item $\Theta(n,{\cal H}) \equiv Sp(n)$ transitive on $S^{(4n-1)}$.
\end{romanlist}
Since we may regard our pure state as a normalized vector in ${\cal C}^{N}$ and thus 
as a point on $S^{(2N-1)}$, we obtain pure state controllability only for $U(N)$ (or 
$SU(N)$ if we are not too fussy about phases) and $Sp(N/2)$, the latter for even $N$ 
only. (Note that we cannot get $O(2N)$ as a subalgebra of $U(N)$.)

Complete controllability is clearly a stronger condition than pure state 
controllability.  To illustrate our theme of the limitations on quantum control, we 
now give two  examples based on a truncated oscillator with nearest-level interactions
for which the algebras generated are $so(N)$ and $sp(N/2)$.  Both these examples are 
generic.

\section{Examples}
\subsection{Three-level oscillator with dipole interactions.}

Consider a three-level system with energy levels $E_1$, $E_2$, $E_3$ and assume the
interaction with an external field $f_1$ is of dipole form with nearest neighbor
interactions only.  Then we have ${H}={H}_0+f(t){H}_1$, where the matrix
representations of ${H}_0$ and $ {H}_1$  are
\[
 {H}_0 =
 \left[\begin{array}{ccc} E_1 & 0 & 0 \\ 0 & E_2 & 0 \\ 0 & 0 & E_3 \end{array}\right], 
 \quad
 {H}_1 =
 \left[\begin{array}{ccc} 0 & d_1 & 0 \\ d_1 & 0 & d_2 \\ 0 & d_2 & 0 \end{array}\right].
\]
If the energy levels are equally spaced, i.e., $E_2-E_1=E_3-E_2=\mu$ and the transition
dipole moments are equal, i.e., $d_1=d_2=d$ then we have
\[
 {H}_0' = \mu
  \left[\begin{array}{ccc}
   -1 & 0 & 0 \\ 0 & 0 & 0 \\ 0 & 0 & +1 \end{array}\right],
  \quad
   {H}_1 = d
  \left[\begin{array}{ccc} 0 & 1 & 0 \\ 1 & 0 & 1 \\ 0 & 1 & 0 \end{array}\right]
\]
where ${H}_0'$ is the traceless part of ${H}_0$.  Both ${\rm i}{H}_0'$ and 
${\rm i}{H}_1$ satisfy
\[
  A+A^{\dagger}=0 \qquad  AJ+J\tilde{A}=0
\]
where
\[
  J =
  \left[\begin{array}{ccc} 0 & 0 & 1 \\ 0 & -1 & 0 \\ 1 & 0 & 0 \end{array}\right]
\]
which is a defining relation for $so(3)$.  The dynamical Lie algebra in this case is 
in fact $so(3)$.  It is easy to show that the matrix $B=UAU^{\dagger}$ is a real 
anti-symmetric representation of $so(3)$ if $U=U^{*}J$.  Explicitly, a suitable unitary
matrix is given by
\[
  U \equiv
  \left [\begin {array}{ccc} 1/\sqrt{2} & 0 & 1/\sqrt{2}
 \\\noalign{\medskip}
                {\rm i}/\sqrt{2} & 0 & -{\rm i}/\sqrt{
2} \\\noalign{\medskip}
		0 & {\rm i} & 0
	\end {array}\right ].
\]
Since the dynamical algebra and group in the basis determined by $U$ consists of 
{\em real} matrices, real states can only be transformed to real states;  this means
that for any initial state there is a large class of unreachable states.  This example
is generic as it applies to $N$-level systems, although for even $N$ we need other 
than dipole interactions to generate $so(N)$.  The analogous dipole interaction 
generates $sp(N/2)$ in the even $N$ case, as we now illustrate.

\subsection{Four-level oscillator with dipole interactions.}

Consider a four-level system with Hamiltonian ${H}={H}_0+f(t){H}_1$,
\[
   {H}_0 = \left( \begin{array}{cccc} 
                    -E_1 &    0 &   0  &   0 \\
                       0 & -E_2 &   0  &   0 \\
                       0 &    0 & +E_2 &   0 \\
                       0 &    0 &   0  & +E_1 
                \end{array} \right),
\]
and
\[
   {H}_1 = \left( \begin{array}{cccc} 
                     0   & +d_1 & 0    &  0   \\
                    +d_1 & 0    & +d_2 &  0   \\
                     0   & +d_2 & 0    & -d_1 \\
                     0   & 0    & -d_1 &  0   
               \end{array} \right).
\]
Note that ${\rm i} {H}_0$ and ${\rm i} {H}_1$ satisfy
\begin{equation} \label{eq:sp}
   {x}=- {x}^\dagger, \quad  {x}^T  {J}+ {J} {x}=0
\end{equation}
for
\[
   {J} = \left( \begin{array}{cccc}
                     0 &  0 & 0  & +1 \\
                     0 &  0 & +1 &  0 \\
                     0 & -1 & 0  &  0 \\
                    -1 &  0 & 0  &  0  
               \end{array} \right),
\]
where ${J}$ is unitarily equivalent to
\begin{equation} \label{eq:J}
        \left( \begin{array}{c|c} 0 &  {I}_{N/2} \\\hline
                                     - {I}_{N/2} & 0 
               \end{array} \right),
\end{equation}
which is a defining relation for $sp(N/2)$.  Consider an initial state of the form
\begin{equation} \label{eq:rho_0}
    {\rho}_0 = {\rm i} {x}+\alpha  {I}_{N},
\end{equation}
where ${x}$ satisfies (\ref{eq:sp}), it can only evolve into states
\begin{equation} \label{eq:rho_1}
    {\rho}_1 = {\rm i}  {y} + \alpha  {I}_{N},
\end{equation}
where ${y}$ satisfies (\ref{eq:sp}), under the action of a unitary evolution operator
in exponential image of $L$.  Hence, any target state that is not of the form 
(\ref{eq:rho_1}) is not accessible from the initial state (\ref{eq:rho_0}).  Note that
the initial state
\[
    {\rho}_0 = \left( \begin{array}{cccc}
                         0.35 & 0    & 0    & 0 \\
                         0    & 0.30 & 0    & 0 \\
                         0    & 0    & 0.20 & 0 \\
                         0    & 0    & 0    & 0.15
                        \end{array} \right)
\]
is of the form  ${\rho}_0= {x}+ 0.25 {I}_4$ and that
\[
  {\rm i} {x} = {\rm i}\left( \begin{array}{cccc}
                         0.10 & 0    & 0     & 0 \\
                         0    & 0.05 & 0     & 0 \\
                         0    & 0    & -0.05 & 0 \\
                         0    & 0    & 0     & -0.10
                        \end{array} \right)
\]
satisfies (\ref{eq:sp}).  Consider the target state
\[
   {\rho}_1 =  \left( \begin{array}{cccc}
                         0.30 & 0    & 0    & 0 \\
                         0    & 0.35 & 0    & 0 \\
                         0    & 0    & 0.20 & 0 \\
                         0    & 0    & 0    & 0.15
                        \end{array} \right)
\]
which is clearly kinematically admissible since
\[
   {\rho}_1 = \left( \begin{array}{cccc}
                     0 & 1 & 0 & 0 \\
                     1 & 0 & 0 & 0 \\
                     0 & 0 & 1 & 0 \\
                     0 & 0 & 0 & 1 
                \end{array} \right)
                 {\rho}_0
                \left( \begin{array}{cccc}
                     0 & 1 & 0 & 0 \\
                     1 & 0 & 0 & 0 \\
                     0 & 0 & 1 & 0 \\
                     0 & 0 & 0 & 1  
                \end{array} \right)^\dagger
\]
but note that ${\rho}_1= {y}+ 0.25 {I}_4$ and
\[
  {\rm i} {y} = {\rm i}\left( \begin{array}{cccc}
                         0.05 & 0    & 0     & 0 \\
                         0    & 0.10 & 0     & 0 \\
                         0    & 0    & -0.05 & 0 \\
                         0    & 0    & 0     & -0.10
                        \end{array} \right)
\]
does not satisfy (\ref{eq:sp}).  Hence, ${\rho}_1$ is not dynamically accessible
from ${\rho}_0$ for this system.

Given a target state ${\rho}_1$ that is not dynamically accessible from an initial
state ${\rho}_0$, we can easily construct observables whose kinematical upper bound
for its expectation value can not be reached dynamically.  Simply consider ${A}=
{\rho}_1$.  The expectation value of ${A}$ assumes its kinematical maximum only 
when the system is in state ${\rho}_1$.  Since ${\rho}_1$ is not reachable, the 
kinematical upper bound is not dynamically realizable.

\section{Conclusions}

In this short introduction to Quantum Control theory, we have described the 
goals of the subject briefly, and then illustrated the limitations by generic
examples where complete control is not possible.  The tools we have used are, 
in the main, those of classical Lie group theory.  Theoretical problems that 
remain to be tackled include to what extent these non-controllable systems can,
in fact, be controlled; and, of course, the paramount problem of implementing
these controls in practice.

\section*{Acknowledgements}

It gives us great pleasure to acknowledge many conversations with colleagues, 
including  Drs. Hongchen Fu and Gabriel Turinici.  A.I.S would also like to 
thank the Laboratoire de Physique Th\'eorique des Liquides, of the University
of Paris VI, for hospitality during the writing of this note.


\begin{thebibliography}{0}
\bibitem{us1}
H.~Fu, S.~G. Schirmer, and A.~I. Solomon.
\emph{Complete controllability of finite-level quantum systems.}
J. Phys. A {\bf 34}, 1679 (2001)

\bibitem{us2}
S.~G. Schirmer, H.~Fu, and A.~I. Solomon.
\emph{Complete controllability of quantum systems.} 
Phys. Rev. A {\bf 63}, 063410 (2001)

\bibitem{mont}
D. Montgomery and H. Samelson, 
\emph{Transformation Group of Spheres.}
Ann. Math. {\bf 44}(3), 454 (1943)

\end{thebibliography}
\end{document}